\documentclass[12pt]{article}
\usepackage[utf8]{inputenc}
\usepackage[T1]{fontenc}
\usepackage[a4paper,margin=2.5cm]{geometry}
\usepackage{amsmath,amssymb}
\usepackage{physics}
\usepackage{graphicx}
\usepackage{cite}
\usepackage{hyperref}
\usepackage{tikz}
\usetikzlibrary{arrows.meta}
\usepackage{authblk}
\usepackage{mathtools}
\usepackage{etoolbox}

\numberwithin{equation}{section}
\AtBeginDocument{%
  \let\bracksOpen\[%
  \let\bracksClose\]%
  \renewcommand{\[}{\begin{equation}}%
  \renewcommand{\]}{\end{equation}}%
}

\title{On the Schrödinger and Carroll--Schrödinger Equations: Dualities and Applications}

\author[1]{José Rojas}
\author[1]{Enrique Casanova}
\author[1,2]{Melvin Arias}

\affil[1]{\textit{\small Instituto de Física, Universidad Autónoma de Santo Domingo, Av. Alma Mater, Santo Domingo 10105, Dominican Republic}}
\affil[2]{\textit{\small Laboratorio de Nanotecnología, Área de Ciencias Básicas y Ambientales, Instituto Tecnológico de Santo Domingo, Av. Los Próceres, Santo Domingo 10602, Dominican Republic}}

\date{\today}

\begin{document}

\maketitle

\begin{abstract}
We investigate precise structural relations between the standard Schrödinger equation and its Carrollian analogue—the Carroll--Schrödinger equation—in $1{+}1$ dimensions, with emphasis on \emph{dualities}, potential maps, and solution behavior. Our contributions proceed in the order of the paper:
(\textit{i}) we encode both dynamics with operators $\hat{\mathcal H}$ and $\hat{\mathcal F}$ under external potentials and explore conditions for obtaining the same type of solutions within both formalisms; (\textit{ii}) we construct a potential–dependent reparametrization $x=\delta(t)$ mapping the space–independent Carroll equation to the time–independent Schrödinger equation, and derive a Schwarzian relation that specifies the map $\delta$ for any static $V_{\rm sch}$ (with harmonic, Coulomb–like, and free examples);
(\textit{iii}) we relate conserved densities and currents by removing $V_{car}$ through a gauge transform followed by a coordinate inversion, establishing equivalence of the continuity equations;
(\textit{iv}) we obtain a Carrollian dispersion relation from an ultra–boost of the energy–momentum two–vector and also derive the classical limit of the Carroll wave equation via the Hamilton--Jacobi formalism;
(\textit{v}) we place Carroll dynamics on an equal–$x$ Hilbert space $L^2(\mathbb R_t)$, prove unitary $x$–evolution, and illustrate dynamics with an exactly solvable Gaussian packet and finite–time quantization for time–localized perturbations; and
(\textit{vi}) for general $V(x,t)$ we perform a gauge reduction to an \emph{interaction momentum} and set up a controlled Dyson expansion about solvable time profiles.
\end{abstract}

\section{Introduction}

The Schrödinger equation admits several complementary derivations; one standard route is a symmetry construction using (projective/unitary) realizations of Galilean or Schrödinger invariance~\cite{Bargmann1954,LevyLeblond1963,Niederer1972}, alongside canonical (e.g.\ Dirac; Sakurai--Napolitano)~\cite{Dirac1958,SakuraiNapolitano2017}, spectral/operator
–theoretic~\cite{Teschl2014, Stone1932}, and path–integral approaches~\cite{Feynman1948,FeynmanHibbs1965}. At the ultra–relativistic corner, the \emph{Carroll group}—the $c\!\to\!0$ contraction of the Poincaré group—organizes dynamics where space and time exchange kinematic roles. The Carrollian limit, emphasized early on by L\'evy–Leblond~\cite{LevyLeblond1965}, now permeates gravity, holography, and dark energy (e.g.~\cite{Dautcourt1998,Hartong2015,Bergshoeff2017,HenneauxSalgado2021,Herfray2022,HansenObersOlingSoegaard2022,Perez2021,Perez2022,DuvalGibbonsHorvathy2014,Ciambelli2018JHEP,DonnayPRL2022,DonnayPRD2023,BagchiEtAl2016,BagchiCaseForCarroll2024,CiambelliCQG2018,CiambelliMarteau2019,FreidelJaiCQG2023,FreidelJaiJHEP2024,BagchiPRL2023,ArmasPRL2024,deBoerEtAl2022}). On the quantum side, a concrete \emph{Carroll--Schrödinger} wave equation serves as an ultra–relativistic analogue of familiar nonrelativistic equations~\cite{Najafizadeh2025,NajafizadehHAL2024}:
\begin{equation}
i\hbar c\,\partial_x \Psi(x,t)\;+\;\frac{\hbar^2}{2m c^2}\,\partial_t^2 \Psi(x,t)\;=\;0,
\label{eq:carroll_free}
\end{equation}
which is first order in space and second order in time.

This work develops a unified $1{+}1$–dimensional framework that places the Schrödinger and Carroll--Schrödinger equations on the same operator footing and tracks how potentials, solutions, currents, inner products, interactions, and classical reductions translate across the two formalisms, in the order presented in the paper. We begin by encoding the dynamics with differential operators $\hat{\mathcal H}$ (Schrödinger) and $\hat{\mathcal F}$ (Carroll) and identify a ``same–type solutions'' criterion (namely, that the constraint class $\ker\hat{\mathcal F}$ is preserved by the Schrödinger evolution and vice versa) as genuine commutation $[\hat{\mathcal H},\hat{\mathcal F}]=0$. This constrains the relationship between the spatiotemporal Schrödinger potential $V_{sch}(x,t)$ and the Carrollian spatiotemporal potential $V_{car}(x,t)$; we also contrast this with a stricter notion based on identical separated solutions derived from purely time–independent potentials. Next, we construct a potential–dependent reparametrization $x=\delta(t)$ that maps the space–independent Carroll problem to the time–independent Schrödinger equation and derive closed formulas for the dual $V_{\rm sch}(x)$ in terms of $V_{\rm car}(t)$. Eliminating $V_{\rm car}$ yields a Schwarzian relationship that we invert explicitly to specify $\delta$ for any static $V_{\rm sch}$; harmonic, Coulomb–like, and free examples illustrate the dictionary.

Conserved structures are then related by a gauge removal of $ V_{car}$ and a coordinate inversion $(x,ct)\mapsto(ct,x)$, which identifies the Carrollian continuity equation with the Schrödinger one and clarifies the role–reversal of space and time. We connect to relativistic and classical structures: we derive a Carrollian dispersion relation by ultra–boosting the energy–momentum two–vector, providing a direct route from Lorentz kinematics to the Carroll sector. We also work out the classical limit of the Carroll--Schrödinger dynamics via the Hamilton--Jacobi formalism and explore the fundamental differences with the Newtonian analog.

We formulate Carroll dynamics on the equal–$x$ Hilbert space $L^{2}(\mathbb{R}_t)$, identify the self-adjoint $x$-evolution generator and its domain, and verify unitarity. We also explore the dynamics of Carroll--Schrödinger solutions by analyzing exactly solvable models, including a closed-form Gaussian packet and \emph{finite–time quantization} for time-localized interactions. For general $V(x,t)$, a unitary gauge reduction produces an \emph{interaction momentum}
$F(x,t)=\,\partial_x\!\int^{t}\!V(x,\tau)\,d\tau$ and recasts the problem as Schrödinger–type evolution in $x$ with the minimal substitution $p\mapsto p-F$; we develop a controlled Dyson expansion in the spatial evolution variable and discuss the structural similarity of the gauge-transformed equation to the equations describing the evolution of temporal solitons and Cherenkov radiation in soliton media \cite{AgrawalNFO2019, ZhengLiu2022}.

\medskip
\noindent\textit{Guide to the paper.}
Section~\ref{section1} formulates the shared–solutions conditions and fixes the admissible potentials. Section~\ref{section2} builds the map $x=\delta(t)$, derives the Schwarzian relation, and presents solvable examples. Section~\ref{section4} relates currents and densities between both formalisms. Sections~\ref{section5} and~\ref{section5'} derive the Carrollian energy–momentum relation via an ultra–boost and obtain the Hamilton--Jacobi classical limit from the Carroll--Schrödinger equation. Section \ref{section6} introduces the Carrollian inner product and establishes the equal–$x$ Hilbert–space picture. Section \ref{section7} analyzes the behavior of exact solutions of the Carroll--Schrödinger equation, outlining two important cases (Gaussian correspondence and finite–time quantization). It also outlines a general solution method for a given potential $V_{car}(x,t)$ via a unitary gauge reduction of the equation and the introduction of an effective interaction–momentum and a spatial Dyson expansion, with physical interpretations.

\section{Shared Solutions Conditions}\label{section1}

Merging the Schrödinger and Carroll--Schrödinger equations with a potential,
\[
-\frac{\hbar^2}{2m}\,\frac{\partial^2 \Psi}{\partial x^2} + V_{sch}(x,t)\,\Psi - i\hbar\,\frac{\partial \Psi}{\partial t} = 0,
\label{eq:sch_potential}
\]
\[
i\hbar c\,\frac{\partial \Psi}{\partial x} \;-\; \frac{1}{2m c^2}\,\Big(-i\hbar\,\frac{\partial}{\partial t} - V_{car}(x,t)\Big)^{2}\Psi = 0.
\label{eq:carroll_potential}
\]

Defining the operators
\[
\hat{\mathcal{H}}=\frac{\hat p^{\,2}}{2m}+\hat V_{sch}(x,t)-\hat E,
\qquad
\hat{\mathcal{F}}=c\,\hat{\tilde p}-\frac{\big(\hat{\tilde E}-\hat{V}_{car}(x,t)\big)^2}{2mc^2},
\label{eq:operators_definition}
\]
where \(\hat p=-i\hbar\,\partial_x\) and \(\hat E= i\hbar\,\partial_t\) are the usual Schr\"odinger generators, while the Carroll generators are (see Sec.~\ref{section6})
\begin{equation}
\hat{\tilde p}=+\,i\hbar\,\frac{\partial}{\partial x},
\qquad
\hat{\tilde E}=-\,i\hbar\,\frac{\partial}{\partial t}.
\label{eq:tilde_ops}
\end{equation}

With these definitions, Eqs.~\eqref{eq:sch_potential}–\eqref{eq:carroll_potential} read
\[
\hat{\mathcal H}\Psi=0,\qquad \hat{\mathcal F}\Psi=0.
\label{eq:H_and_F_zero}
\]

We ask for a relation between the potentials that makes the two equations
compatible in the sense that each operator preserves the constraint
subspace defined by the other. A standard sufficient condition is vanishing commutator on a common
invariant core $\mathcal D$,
\begin{equation}
[\hat{\mathcal H},\hat{\mathcal F}]=0 \quad \text{on } \mathcal D.
\label{eq:lax_form}
\end{equation}

If $\psi\in\mathcal D$ and $\hat{\mathcal F}\psi=0$, then
$\hat{\mathcal F}(\hat{\mathcal H}\psi)=\hat{\mathcal H}(\hat{\mathcal F}\psi)=0$,
so $\hat{\mathcal H}$ maps $\ker\hat{\mathcal F}\cap\mathcal D$ into itself; the
same argument with $\hat{\mathcal H}$ and $\hat{\mathcal F}$ swapped shows that
$\hat{\mathcal F}$ preserves $\ker\hat{\mathcal H}\cap\mathcal D$. Thus the two
equations admit a stable class of \emph{shared solutions} (but this does not
imply $\ker\hat{\mathcal H}=\ker\hat{\mathcal F}$).

For the operators above, a direct computation of the commutator shows that the only way \eqref{eq:lax_form} can hold is if and only if
\[
\partial_x V_{sch}=\partial_x V_{car}=0,\qquad \partial_t V_{sch}=\,\partial_t V_{car},
\]
i.e.
\[
\boxed{\; V_{sch}(x,t)=V(t),\qquad   V_{car}(x,t)=\,V(t)+C \;(C\in\mathbb R)\; }.
\label{eq:compat_condition}
\]
Under \eqref{eq:compat_condition} we have $[\hat{\mathcal H},\hat{\mathcal F}]=0$ on a common invariant core, so $\hat{\mathcal H}$ maps $\ker\hat{\mathcal F}$ into itself and vice versa.

\medskip
\noindent\emph{Remark on a stronger requirement and the opposite sign.}
One may instead ask that \emph{every} separated Schrödinger solution for $V_{sch}=V_{sch}(t)$ also solve the Carroll equation with the \emph{same} wavefunction.
Let $\Psi(x,t)=\phi_k(x)T_k(t)$ with
\[
-\frac{\hbar^2}{2m}\phi_k''=E_k\phi_k,\qquad
i\hbar\,\dot T_k=(E_k+V_{sch}(t))\,T_k.
\]
Substituting this into the Carroll operator and using $\hat{\tilde E}=-i\hbar\partial_t$,
\[
[\hat{\tilde E}-\hat{V}_{car}(t)]\Psi
=-\,[E_k+V_{sch}(t)+ V_{car}(t)]\Psi,
\quad\Rightarrow\quad
[\hat{\tilde E}-\hat{V}_{car}(t)]^{2}\Psi=[E_k+V_{sch} +V_{car}]^{2}\Psi.
\]
Then $\hat{\mathcal F}\Psi=0$ becomes
\[
i\hbar c\,\frac{\phi_k'}{\phi_k}
=\frac{[E_k+V_{sch}(t)+ V_{car}(t)]^{2}}{2mc^2}.
\]
The left side depends only on $x$, the right side only on $t$, so both must be time–independent; hence
\[
\boxed{\  V_{car}(t)=-\,V_{sch}(t)+C\ }.
\]
It is important to note that this stronger notion (different from the commutation/invariance used above) flips the sign relative to \eqref{eq:compat_condition}.

\section{From the Time-Independent Schrödinger Equation to the Space-Independent Carroll--Schrödinger Equation}\label{section2}

Analogous to the derivation of the time–independent Schrödinger equation, a \emph{space–independent} Carroll equation follows from \eqref{eq:carroll_potential} when $V_{car}(x,t)= V_{car}(t)$.
Dividing \eqref{eq:carroll_potential} by $c$ and seeking separated states $\Psi(x,t)=\phi_x(x)\,\phi_t(t)$ gives
\begin{equation}
i\hbar\,\frac{\phi_x'}{\phi_x}
=\frac{1}{2mc^{3}}\,
\frac{\big(-i\hbar\,\partial_t- V_{car}(t)\big)^{2}\phi_t}{\phi_t}
=:p_{0},
\qquad p_{0}\in\mathbb R,
\label{eq:car_sep_const}
\end{equation}
so that the spatial factor is a momentum eigenstate,
\begin{equation}
\phi_x(x)=\exp\!\Big(-\frac{i}{\hbar}p_{0}x\Big),
\qquad \hat{\tilde p}\,\phi_x=p_{0}\,\phi_x,
\label{eq:car_phix}
\end{equation}
and the time factor satisfies the \emph{space–independent Carroll equation}
\begin{equation}
\frac{1}{2mc^{3}}\Big(-i\hbar\,\frac{d}{dt}- V_{car}(t)\Big)^{2}\phi_t(t)-p_{0}\,\phi_t(t)=0.
\label{eq:car_time_p0}
\end{equation}
Multiplying \eqref{eq:car_time_p0} by $c$ one obtains the energy form
\begin{equation}
\boxed{\;\frac{1}{2mc^{2}}\Big(-i\hbar\,\frac{d}{dt}- V_{car}(t)\Big)^{2}\phi_t(t)-cp_{0}\,\phi_t(t)=0.\;}
\label{eq:car_time_E0}
\end{equation}
Remembering that
\begin{equation}
cp_{0}=\frac{E_{0}^{2}}{2mc^{2}},
\qquad\text{so that}\quad E_{0}=\pm\sqrt{2mc^{3}p_{0}},
\label{eq:p0E0}
\end{equation}
we may observe that the \emph{defining} separation parameter is the Carroll momentum $p_{0}$ (appearing linearly in \eqref{eq:car_sep_const}–\eqref{eq:car_time_p0}); the energy label $E_{0}$ is not unique since $E_{0}$ and $-E_{0}$ correspond to the same $p_{0}$ via \eqref{eq:p0E0}.

Having just derived the space–independent Carroll--Schrödinger equation \ref{eq:car_time_E0}, is it possible to find a coordinate transformation that will take us to the time–independent Schrödinger equation or vice versa?
\[
-\frac{\hbar^2}{2m} \frac{\partial^2 \Psi_x}{\partial x^2} + {V_{sch}}(x) \Psi_x(x) = E_{sch} \Psi_x(x)
\label{eq:sch_time_indep}
\]
Exploring a general case where $x$ is an arbitrary function of $t$, we have 
\[
x = \delta(t), \quad \Psi(x) \rightarrow \Psi(\delta(t)) = \Phi(t)
\label{eq:map_delta}
\]
the time–independent Schrödinger equation becomes:
\[
-\frac{\hbar^2}{2m} \left[\frac{\ddot{\Phi}}{\dot{\delta}^2} - \frac{\dot{\Phi} \ddot{\delta}}{\dot{\delta}^3}\right] + {V}_{sch}(\delta(t)) \Phi - E_{sch} \Phi = 0
\label{eq:sch_transformed}
\]
Multiplying Eq.~\ref{eq:sch_transformed} by $ \dot{\delta}^2/c^2 $ and comparing with the expanded Carroll--Schrödinger equation, taking $\Psi(t)=\Phi(t)$, 
\[
-\frac{\hbar^2}{2m c^2} \frac{\partial^2 \Psi}{\partial t^2} + \frac{i\hbar {V}_{car} \dot{\Psi}}{m c^2} + \left[ \frac{i\hbar}{2m c^2}      \frac{\partial {V}_{car}}{\partial t} + \frac{{V}^2 _{car}  }{2m c^2}    - cp_0\right]\Psi = 0
\label{eq:car_expand3}
\]
we obtain 
\[
\frac{\ddot{\delta}}{\dot{\delta}} = \frac{2i}{\hbar}V_{car}(t)
\;\Rightarrow\;
\delta(t) = C_1 + C_0 \int^{t} \exp\!\left(\frac{2i}{\hbar}\int^{t'} V_{car}(s)\,ds\right) dt' .
\]
(with any lower integration limits properly absorbed into the constants $C_0$ and $C_1$) and,
\[
V_{sch}(x) = E_{sch} + \frac{1}{2m(\dot{\delta})^2}\left[i\hbar \frac{dV_{car}(t)}{dt} + V_{car}^2(t) - E_{0}^2\right]_{t=\delta^{-1}(x)}
\label{eq:Vsch_from_Vcar}
\]
This is the transformation rule between the systems. Notably, the reparametrization \(x=\delta(t)\) is potential dependent. We have found that, given a space–independent Carroll--Schrödinger equation with a specified potential, it may be transformed into a space–dependent Schrödinger equation under the map
\[
t = \delta^{-1}(x), \quad \text{where} \quad x = \delta(t) = C_1 + C_0 \int^{t} \exp\!\left(\frac{2i}{\hbar}\int^{t'} V_{car}(s)\,ds\right) dt'
\label{eq:inverse_map}
\]
\[
V_{sch}(x) = E_{sch} + \frac{1}{2m(\dot{\delta})^2}\left[i\hbar \frac{dV_{car}(t)}{dt} + V_{car}^2(t) - E_{0}^2\right]_{t=\delta^{-1}(x)}
\label{eq:Vsch_from_Vcar_again}
\]
so that the solution of the Carroll differential operator $  \Phi_{car}(t)  $ equals the solution of the equivalent Schrödinger operator $ \Psi (x) $ evaluated at $ x = \delta(t)$.

Alternatively, given a velocity profile of a given parametrization $v(t)=\dot{x}(t)=\dot{\delta}(t)$, an associated Carroll potential is
\[
V_{car}(t)= -\frac{i\hbar}{2} \frac{\dot{v}(t)}{v(t)}, \quad \text{and an associated dual Schrödinger potential is}
\label{eq:Vcar_from_v}
\]
\[
V_{sch}(x) = 
E_{sch} + \frac{1}{2m v^2}\left[
\frac{\hbar^2}{2} \frac{d}{dt}\left(\frac{\dot{v}(t)}{v(t)}\right)
- \frac{\hbar^2}{4}\left(\frac{\dot{v}(t)}{v(t)}\right)^2
 - E_{0}^2
\right]_{t=\delta^{-1}(x)}
\label{eq:Vsch_from_v}
\]

\noindent\textbf{Potential inversion procedure.}
We begin from the reparametrization relations (with $x=\delta(t)$)
\begin{equation}
\frac{\ddot\delta}{\dot\delta}=\frac{2i}{\hbar}V_{\mathrm{car}}(t),
\qquad
V_{\mathrm{sch}}(\delta(t))
=E_{\mathrm{sch}}+\frac{1}{2m\,\dot\delta(t)^{2}}\Big(i\hbar\,\dot V_{\mathrm{car}}(t)+V_{\mathrm{car}}(t)^{2}-E_{0}^{2}\Big),
\label{eq:map_start}
\end{equation}
Eliminating $V_{\mathrm{car}}$ via $V_{\mathrm{car}}= -\frac{i\hbar}{2}\frac{\ddot\delta}{\dot\delta}$ gives
\begin{equation}
V_{\mathrm{sch}}(\delta(t))-E_{\mathrm{sch}}
=\frac{\hbar^{2}}{4m}\,\frac{\{\delta,t\}}{\dot\delta(t)^{2}}
-\frac{E_{0}^{2}}{2m}\,\frac{1}{\dot\delta(t)^{2}},
\qquad
\{\delta,t\}:=\frac{\delta'''}{\delta'}-\frac{3}{2}\!\left(\frac{\delta''}{\delta'}\right)^{2}.
\label{eq:forward_S}
\end{equation}
where  $ \{\delta,t\} $ is the Schwarzian derivative \cite{Schwarz1873,NehariConformal}. Writing $\tau(x):=\delta^{-1}(x)$ and using the inversion identities
\(
\big(\{\delta,t\}/\dot\delta^{2}\big)\!\big|_{t=\tau(x)}=-\{\tau,x\},\;
\big(1/\dot\delta^{2}\big)\!\big|_{t=\tau(x)}=\tau'(x)^{2},
\)
we obtain the inverse equation
\begin{equation}
\boxed{\;
\{\tau,x\}+\frac{2E_{0}^{2}}{\hbar^{2}}\,\tau'(x)^{2}
=-\,\frac{4m}{\hbar^{2}}\Big(V_{\mathrm{sch}}(x)-E_{\mathrm{sch}}\Big).
\;}
\label{eq:inverse_master_narr}
\end{equation}
To linearize, use the Schwarzian chain rule $\{f\!\circ\!\tau,x\}=\{f,\tau\}\tau'^{2}+\{\tau,x\}$ and pick $f$ with constant Schwarzian $\{f,u\}=2(E_{0}/\hbar)^{2}$; an explicit choice is $f(u)=\tan\!\big(\tfrac{E_{0}}{\hbar}u\big)$. Define
\[
\sigma(x):=f(\tau(x))=\tan\!\Big(\tfrac{E_{0}}{\hbar}\,\tau(x)\Big),
\]
so that \(\{\sigma,x\}=\{\tau,x\}+\tfrac{2E_{0}^{2}}{\hbar^{2}}\tau'^{2}\), and \eqref{eq:inverse_master_narr} becomes the pure Schwarzian equation
\begin{equation}
\boxed{\;\{\sigma,x\}=-\,\frac{4m}{\hbar^{2}}\Big(V_{\mathrm{sch}}(x)-E_{\mathrm{sch}}\Big).\;}
\label{eq:pure_S_narr}
\end{equation}
A standard construction then solves \eqref{eq:pure_S_narr}: if $y_{1},y_{2}$ form a fundamental system of
\begin{equation}
y''(x)+q(x)\,y(x)=0,
\qquad
q(x):=\frac{2m}{\hbar^{2}}\Big(V_{\mathrm{sch}}(x)-E_{\mathrm{sch}}\Big),
\label{eq:lin_ode_q_narr}
\end{equation}
then $\sigma(x)=y_{1}(x)/y_{2}(x)$ satisfies $\{\sigma,x\}=-2q(x)$, hence \eqref{eq:pure_S_narr}. Therefore
\begin{equation}
\sigma(x)=\frac{y_{1}(x)}{y_{2}(x)},
\qquad
\tau(x)=\frac{\hbar}{E_{0}}\arctan\!\big(\sigma(x)\big),
\qquad
\boxed{\,\delta(t)=\tau^{-1}(t)\,}
\label{eq:delta_construct_narr}
\end{equation}
(on any interval where $\tau$ is monotone). Substituting \eqref{eq:delta_construct_narr} into \eqref{eq:inverse_master_narr} and inverting reproduces \eqref{eq:forward_S}, hence \eqref{eq:map_start}; this verifies that the constructed $\delta$ indeed yields precisely the target $V_{\mathrm{sch}}$ for the chosen $E_{0}$ (equivalently $p_{0}$).

\medskip
\noindent\textbf{Specific targets.} 
We now record the corresponding linear ODE \eqref{eq:lin_ode_q_narr}, a convenient fundamental pair $(y_{1},y_{2})$, and the resulting $\delta$ for three canonical static potentials.

\smallskip
\emph{Harmonic oscillator:} \(V_{\mathrm{sch}}(x)=\tfrac{1}{2}m\omega^{2}(x-x_{0})^{2}\).
With \(\xi=\sqrt{\tfrac{m\omega}{\hbar}}(x-x_{0})\), \(z=\sqrt{2}\,\xi\), and \(\nu=\tfrac{E_{\mathrm{sch}}}{\hbar\omega}-\tfrac{1}{2}\),
the ODE \eqref{eq:lin_ode_q_narr} is the parabolic–cylinder equation
\[
\frac{d^{2}y}{dz^{2}}+\Big(\nu+\tfrac{1}{2}-\frac{z^{2}}{4}\Big)y=0.
\]
A fundamental pair is \(y_{1}(x)=D_{\nu}(z)\), \(y_{2}(x)=D_{\nu}(-z)\). Hence
\[
\sigma(x)=\frac{D_{\nu}\!\big(\sqrt{2}\,\xi\big)}{D_{\nu}\!\big(-\sqrt{2}\,\xi\big)},\qquad
\tau(x)=\frac{\hbar}{E_{0}}\arctan\!\big(\sigma(x)\big),\qquad
\delta(t)=\tau^{-1}(t),
\]
which, by the construction above, produces \emph{exactly} \(V_{\mathrm{sch}}(x)=\tfrac{1}{2}m\omega^{2}(x-x_{0})^{2}\).

\smallskip
\emph{Coulomb–like:} 
\[
V_{\mathrm{sch}}(x) = \frac{\alpha}{x - x_{0}}
\quad \text{(on an interval } x \gtrless x_{0}\text{).}
\]
Set 
\[
k = \frac{\sqrt{2m|E_{\mathrm{sch}}|}}{\hbar}, 
\quad 
z = 2k(x - x_{0}), 
\quad 
\kappa = \frac{m\alpha}{\hbar^{2}k}.
\]
Then \eqref{eq:lin_ode_q_narr} becomes the Whittaker normal form with 
\(\mu = \tfrac{1}{2}\).

\[
\frac{d^{2}Y}{dz^{2}}+\left(-\frac{1}{4}+\frac{\kappa}{z}+\frac{1/4-\mu^{2}}{z^{2}}\right)Y=0,
\qquad \mu=\frac{1}{2}\;\Rightarrow\;\frac{1}{4}-\mu^{2}=0,
\]
i.e.
\[
Y_{zz}+\left(-\frac{1}{4}+\frac{\kappa}{z}\right)Y=0.
\]
A fundamental pair is \(y_{1}(x)=M_{\kappa,\frac{1}{2}}\!\big(2k(x-x_{0})\big)\), \(y_{2}(x)=W_{\kappa,\frac{1}{2}}\!\big(2k(x-x_{0})\big)\). Thus
\[
\sigma(x)=\frac{M_{\kappa,\frac{1}{2}}\!\big(2k(x-x_{0})\big)}{W_{\kappa,\frac{1}{2}}\!\big(2k(x-x_{0})\big)},
\qquad
\tau(x)=\frac{\hbar}{E_{0}}\arctan\!\big(\sigma(x)\big),
\qquad
\delta(t)=\tau^{-1}(t),
\]
which yields \emph{exactly} \(V_{\mathrm{sch}}(x)=\alpha/(x-x_{0})\).

\smallskip
\emph{Free case:} \(V_{\mathrm{sch}}(x)\equiv 0\).
Taking \(E_{\mathrm{sch}}=0\) gives \(q\equiv 0\), so a fundamental pair is \(y_{1}(x)=1\), \(y_{2}(x)=x-x_{0}\).
Hence \(\sigma(x)=1/(x-x_{0})\), \(\tau(x)=\tfrac{\hbar}{E_{0}}\arctan\!\big(\tfrac{1}{x-x_{0}}\big)\), and \(\delta=\tau^{-1}\),
for which \eqref{eq:inverse_master_narr} (equivalently \eqref{eq:forward_S}) yields \(V_{\mathrm{sch}}\equiv 0\).

\medskip
\noindent\textit{Remarks.} (i) The construction applies on any subinterval where $y_{2}$ has no zeros so that $\tau$ is monotone and invertible; different branches correspond to different coordinate patches. (ii)
Regarding the hermiticity of $V_{\mathrm{car}}$ and the phase choice, from \eqref{eq:map_start} one has
\begin{equation}
V_{\mathrm{car}}(t)
=-\,\frac{i\hbar}{2}\,\frac{\ddot\delta(t)}{\dot\delta(t)}
=-\,\frac{i\hbar}{2}\,u(t),
\qquad
u(t):=\frac{\ddot\delta}{\dot\delta}.
\label{eq:Vcar-Herm-def}
\end{equation}
Requiring a \emph{Hermitian} (real–valued) $V_{\mathrm{car}}$ is therefore equivalent to demanding
\begin{equation}
u(t)\in i\mathbb{R}\quad\Longleftrightarrow\quad
\frac{\ddot\delta}{\dot\delta}\ \text{is purely imaginary}.
\label{eq:Hermiticity-condition}
\end{equation}
Writing $\tau=\delta^{-1}$, the inverse–function relations give
\(
\frac{\ddot\delta}{\dot\delta}=-\frac{\tau''}{\tau'^2}.
\)
Hence a sufficient (and convenient) way to enforce \eqref{eq:Hermiticity-condition} is to pick a branch with
\(
\tau'(x)\in i\mathbb{R}
\)
on the $x$–interval of interest. This is achieved by the constant rescaling (for $|\sigma(x)| \neq 1$)
\begin{equation}
\sigma(x)\ \longrightarrow\ \sigma_H(x):=i\,\sigma(x),
\quad
\tau_H(x):=\frac{\hbar}{E_0}\arctan\big(\sigma_H(x)\big)
= i\,\frac{\hbar}{E_0}\operatorname{artanh}\!\big(\sigma(x)\big),
\quad
\delta_H:=\tau_H^{-1}.
\label{eq:Hermitian-branch}
\end{equation}
Because the Schwarzian is invariant under constant Möbius rescalings,
\(
\{\sigma_H,x\}=\{\sigma,x\},
\)
\eqref{eq:pure_S_narr} and all reconstruction formulas for $V_{\mathrm{sch}}$ remain unchanged. In general, to ensure the hermiticity of the potentials (which would then require a complex map), the transformations already found need only be rescaled by the factor shown above.

\section{Probability Currents and Densities}\label{section4}

It is well known that for the Schrödinger equation the probability density and current are
\[
\rho_{sch} = \Psi_{sch}^* \Psi_{sch} , \quad 
J^x_{sch} = \frac{i\hbar}{2m}\!\left(\Psi_{sch} \frac{\partial \Psi^* _{sch}}{\partial x} - \Psi^* _{sch} \frac{\partial \Psi_{sch}}{\partial x}\right),
\label{eq:sch_current_density}
\]
satisfying the continuity equation $\partial_t \rho + \partial_x J_x = 0$.

For the Carroll--Schrödinger case, a straightforward calculation gives
\[
J^x_{car} = \Psi^* _{car} \Psi _{car},\qquad 
\rho_{car} = \frac{i\hbar}{2m c^3} 
\left( \Psi^* _{car} \frac{\partial \Psi _{car}}{\partial t} - \Psi _{car}\frac{\partial \Psi^* _{car}}{\partial t} \right)
+ \frac{\Psi^* _{car} \Psi _{car} V_{car}}{m c^3}.
\label{eq:car_J_and_rho}
\]
Thus, spatial and temporal roles effectively interchange relative to the ordinary Schrödinger case; a notable feature is that the Carroll density is potential dependent.

Again, we may relate these two results under a certain transformation. In $d=1+1$ with $J^\mu=(\rho,\vec J)$,
\[
J^\mu_{carroll} = 
\left(
\frac{i\hbar}{2m c^3}
\left( \Psi^* _{car} \frac{\partial \Psi _{car}}{\partial t} - \Psi _{car}\frac{\partial \Psi^* _{car}}{\partial t} \right)
+ \frac{\Psi^* _{car}\Psi _{car} V_{car}(x,t)}{m c^3}, \;
\Psi^* _{car} \Psi _{car}
\right),
\label{eq:Jmu_carroll}
\]
\[
J^\mu_{sch} = 
\left(
\Psi^* _{sch}\Psi _{sch}, \;
\frac{i\hbar}{2m}\left[
\Psi _{sch}\frac{\partial \Psi^* _{sch}}{\partial x} - \Psi^* _{sch} \frac{\partial \Psi _{sch}}{\partial x}
\right]
\right).
\label{eq:Jmu_sch}
\]

First, removing the explicit $V_{car}(x,t)$ from \ref{eq:Jmu_carroll} by performing a gauge transformation of the form
\[
\Phi_{car}(x,t)=\exp\!\left[-\frac{i}{\hbar}\int^{t}V_{car}(x,\tau)\,d\tau\right]\Psi_{car}(x,t),
\]
so that
\[
\rho_{car}=\frac{i\hbar}{2mc^{3}}\!\left(\Phi^{*}_{car}\partial_{t}\Phi_{car}-\Phi_{car}\,\partial_{t} \Phi^{*}_{car}\right),
\qquad
J^{x}_{car}=|\Phi_{car}|^{2}.
\]
Now, performing a coordinate inversion of the form
\[
x'=ct,\qquad t'=\,\frac{x}{c},
\]

The Carroll continuity equation \(\partial_{x}\rho_{car}+\partial_{t}J^{x}_{car}=0\) becomes
\[
\partial_{t'}|\Phi|^{2}
+\partial_{x'}\!\left(\frac{i\hbar}{2m}\big[\Phi\,\partial_{x'}\Phi^{*}-\Phi^{*}\partial_{x'}\Phi\big]\right)=0,
\]
which is exactly the Schr\"odinger continuity equation. We observe that the two
continuity equations have the same form, up to a gauge transformation followed
by a coordinate inversion in the
\((x,ct)\) plane. 

\section{From Schrödinger to Carroll--Schrödinger in $d=1{+}1$ via a $V\!\to\!\infty$ boost}\label{section5}

In $d=1{+}1$ the energy–momentum two–vector is $p^\mu=(E/c,\,P)$ and transforms under a standard Lorentz boost with speed $V$ (along $+x$) as
\begin{equation}
\frac{E'}{c}=\gamma\!\left(\frac{E}{c}-\beta P\right),\qquad
P'=\gamma\!\left(P-\beta\,\frac{E}{c}\right),
\qquad
\beta:=\frac{v}{c},\ \ \gamma:=\frac{1}{\sqrt{1-\beta^{2}}}.
\label{eq:LT-1p1}
\end{equation}
To evaluate the formal ultra–boost $V\to\infty$, the Lorentz formulas are analytically continued to $|\beta|>1$ by choosing the branch
\begin{equation}
\gamma=\frac{i}{\sqrt{\beta^{2}-1}}\quad(|\beta|>1),
\qquad \gamma\sim \frac{i}{\beta}\ \ \text{as}\ \ \beta\to\infty.
\label{eq:gamma-branch}
\end{equation}
Substituting \eqref{eq:gamma-branch} into \eqref{eq:LT-1p1} yields the ultra–boosted limits
\begin{equation}
\frac{E'}{c}=\gamma\!\left(\frac{E}{c}-\beta P\right)\xrightarrow[\beta\to\infty]{}-\,i\,P,
\qquad
P'=\gamma\!\left(P-\beta\frac{E}{c}\right)\xrightarrow[\beta\to\infty]{}-\,\frac{i}{c}\,E,
\end{equation}
i.e.
\begin{equation}
\boxed{\,E'=-\,i\,c\,P,\qquad P'=-\,\frac{i}{c}\,E\,}.
\label{eq:UR-map}
\end{equation}
On the chosen branch the invariant $p^\mu p_\mu=(E/c)^2-P^2$ is preserved.

The free Schrödinger energy–momentum relation,
\begin{equation}
E=\frac{P^{2}}{2m}\qquad\Longleftrightarrow\qquad E'=\frac{P'^2}{2m},
\label{eq:S-dispersion}
\end{equation}
combined with \eqref{eq:UR-map} gives
\begin{equation}
\frac{(-iE/c)^2}{2m}=-\,i\,c\,P
\ \Longrightarrow\
-\,\frac{E^{2}}{2m c^{2}}=-\,i\,c\,P
\ \Longrightarrow\
\boxed{\,\frac{E^{2}}{2m c^{3}}=i\,P\,}.
\label{eq:Carroll-dispersion}
\end{equation}
Observing the relation, we notice that the Carroll--Schrödinger energy–momentum relation is recovered under the mass redefinition $m\mapsto i m$, as in~\cite{NajafizadehHAL2024,Najafizadeh2025}.

\section{Hamilton--Jacobi Limit}\label{section5'}

Starting from the free Carroll--Schr\"odinger equation \ref{eq:carroll_free}
\begin{equation}
i\hbar c\,\partial_x \Psi(x,t) + \frac{\hbar^2}{2mc^2}\,\partial_t^2 \Psi(x,t)=0,
\label{eq:fundamental}
\end{equation}
we use the WKB \emph{ansatz}
\begin{equation}
\Psi(x,t)=A(x,t)\,e^{\frac{i}{\hbar}S(x,t)} .
\end{equation}
Substituting into \eqref{eq:fundamental} and grouping real and imaginary terms yields the two equations
\begin{equation}\label{eq:jacobi-real}
\frac{\hbar}{2mc^{3}}\frac{\partial^{2}A}{\partial t^{2}}
-\frac{A}{2mc^{3}\hbar}\Bigl(\frac{\partial S}{\partial t}\Bigr)^{2}
-\frac{A}{\hbar}\frac{\partial S}{\partial x} = 0
\end{equation}
\begin{equation}\label{eq:jacobi-imagin}
\frac{1}{mc^{3}}\frac{\partial A}{\partial t}\frac{\partial S}{\partial t}
+\frac{\partial A}{\partial x} = 0
\end{equation}
Multiplying \ref{eq:jacobi-real} by $\hbar$ and taking the classical limit $\hbar\to 0$ (discarding higher-order terms in $\hbar$), we obtain the associated Hamilton--Jacobi equation
\begin{equation}
\boxed{\;
\frac{1}{2mc^{3}}\Bigl(\frac{\partial S}{\partial t}\Bigr)^{2}
+\frac{\partial S}{\partial x}=0
\;}
\label{eq:HJ}
\end{equation}
for the action $S(x,t)$.

We may solve for the action using a separable ansatz of the form
\[
S(x,t)=S_x(x)+S_t(t),
\]
which yields from the Hamilton--Jacobi equation
\begin{equation}
\frac{1}{2 m c^{3}}\!\left(\frac{\partial S}{\partial t}\right)^{2}
+\left(\frac{\partial S}{\partial x}\right)=0
\quad\Longrightarrow\quad
\frac{1}{2 m c^{3}}\!\left(\frac{dS_t}{dt}\right)^{2}
+\left(\frac{dS_x}{dx}\right)=0 .
\label{eq:HJsep}
\end{equation}
By separation of variables we set
\[
\frac{1}{2 m c^{3}}\!\left(\frac{dS_t}{dt}\right)^{2}=\alpha,
\qquad
\left(\frac{dS_x}{dx}\right)=-\alpha ,
\]
and therefore
\begin{equation}
S(x,t)=-\alpha\,x \pm \sqrt{2 m c^{3}\alpha}\; t + C_{0}.
\label{eq:S-solution}
\end{equation}
Here $\alpha$ is the separation constant with units of momentum; we
relabel it as $p_{0}$ since it is a constant
of motion. (\emph{Sign convention:} The negative sign arises because in our Carroll representation the momentum operator is $\hat p_x=i\hbar\,\partial_x$ (cf.~\cite{Najafizadeh2025}). Acting on a WKB wave $\Psi=A\,e^{iS/\hbar}$ gives $\hat p_x\Psi=(i\hbar\,\partial_xA - A\,\partial_x S)\,e^{iS/\hbar}$; in the eikonal/WKB limit the $i\hbar\,\partial_xA$ term is subleading, so $\hat p_x\Psi\approx(-\partial_x S)\Psi$, hence $p_x=-\partial_x S$).

The second constant follows from
\begin{equation} \label{eq:Jacobi-clas}
\frac{\partial S}{\partial \alpha_i}=\beta_i
\qquad\Longrightarrow\qquad
\frac{\partial S}{\partial p_{0}}
=-x \pm \sqrt{\frac{m c^{3}}{2p_{0}}}\; t=\beta_{\pm}.
\end{equation}
We observe that, from the Carrollian dispersion relation,
\[
E(p)= \pm  \sqrt{2 m c^{3}p}
\quad\Rightarrow\quad
\frac{d {E}}{dp}
= \pm\sqrt{\frac{m c^{3}}{2p}}
=v_{\text{group}} .
\]
Thus, in the classical limit,
\[
v_{\text{particle}}=\pm \sqrt{\frac{m c^{3}}{2p}},
\]
and Eq.~\ref{eq:Jacobi-clas} may be rewritten as 
\begin{equation}   
x=\pm v_{\text{group}}\,t-\beta_{\pm},
\end{equation}
so $\beta_{\pm}$ can be identified with the particle's initial position.

In the classical limit of the one-dimensional free-particle case, the motion is rectilinear and uniform, showing no fundamental difference from the Newtonian limit of the one-dimensional free Schr\"odinger equation. This equivalence is lost once interactions are introduced. Nevertheless, even in the simplest one-dimensional free case, a fundamental difference between the formalisms appears in the momentum–velocity relation. Inverting Eq.~\eqref{eq:Jacobi-clas} and solving for the magnitude of the momentum, we obtain
\begin{equation}\label{exoticmomentum}
v_{p}=\pm \sqrt{\frac{m c^{3}}{2p}}
\qquad\Longrightarrow\qquad
\lvert p\rvert=\frac{m c^{3}}{2 v^{2}}, \quad 
E(p)= \pm  \sqrt{2 m c^{3}p} = \pm \frac{mc^3}{v}
\end{equation}
Consistent with Ref.~\cite{NajafizadehHAL2024}, this result shows an inverse momentum–velocity relationship, in sharp contrast to the Newtonian case.

\subsection*{Extended phase space}

Since time is not the evolution parameter, the extended phase-space approach is natural for analyzing Carrollian classical systems (cf.~\cite{deBoerEtAl2022}). With a Carrollian potential $V_{\rm car}(x,t)$ the Hamilton--Jacobi equation reads 
\begin{equation}
\frac{1}{2mc^{3}}\Bigl(\partial_t S +V_{\rm car}(x,t)\Bigr)^{2}
+\partial_x S=0.
\label{eq:HJ2}
\end{equation}
Identifying the generalized momenta (our convention $\hat p_x=i\hbar\partial_x$ implies $p_x=-\partial_x S$),
\begin{equation}
p_x=-\partial_x S,
\qquad
p_t=\partial_t S,
\label{eq:generalizedmomenta}
\end{equation}
the eikonal/constraint equation is
\begin{equation}
F(x,t;p_x,p_t):=-c\,p_x+\frac{1}{2mc^{2}}\big(p_t + V_{\rm car}(x,t)\big)^{2}=0.
\label{eq:F_master}
\end{equation}
Its characteristic system (dot $=\tfrac{d}{d\lambda}$) is
\begin{equation}\label{eq:parameters-Action}
\dot x=\partial_{p_x}F=-c,\qquad
\dot t=\partial_{p_t}F=\frac{p_t +V_{\rm car}}{mc^{2}},
\end{equation}
\begin{equation}
\dot p_x=-\partial_x F= -\frac{p_t+ V_{\rm car}}{mc^{2}}\partial_x V_{\rm car},
\qquad
\dot p_t=-\partial_t F= -\frac{p_t+ V_{\rm car}}{mc^{2}}\partial_t V_{\rm car}.
\label{eq:char_master}
\end{equation}
From \eqref{eq:parameters-Action} we have $\dot x=-c$. A convenient gauge choice is $\lambda=-x/c$, so that $d/d\lambda=-c\,d/dx$ and $\dot x=-c$ identically. Define
\[
q(x):=p_t(x)+ V_{\rm car}\big(x,t(x)\big).
\]
Then, using gauge-invariant $x$-slopes,
\begin{equation}
\boxed{\quad 
\frac{dt}{dx}=-\,\frac{q(x)}{mc^{3}},\qquad
\frac{dq}{dx}=\,\partial_x V_{\rm car}\big(x,t(x)\big),\qquad
p_x(x)=\,\frac{q(x)^{2}}{2mc^{3}}\quad}
\label{eq:ray_system_q}
\end{equation}
and $t(x)$ obeys
\begin{equation}
\boxed{\ \frac{d^{2}t}{dx^{2}}=-\,\frac{1}{mc^{3}}\,
\partial_x V_{\rm car}\big(x,t(x)\big)\ }.
\label{eq:t_second_order}
\end{equation}

\paragraph{General quadratures.}
For initial data $t(x_0)=t_0$, $q(x_0)=q_0$,
\begin{equation}
q(x)=q_0+\int_{x_0}^{x}\partial_x V_{\rm car}\big(\xi,t(\xi)\big)\,d\xi,\qquad
t(x)=t_0-\frac{1}{mc^{3}}\int_{x_0}^{x} q(\eta)\,d\eta.
\label{eq:general_quadratures}
\end{equation}
This gives the Picard iteration
\begin{align}
t^{(0)}(x)&=t_0-\frac{q_0}{mc^{3}}(x-x_0),\\
q^{(n+1)}(x)&=q_0+\int_{x_0}^{x}\partial_x V_{\rm car}\!\left(\xi,\,t^{(n)}(\xi)\right)d\xi,\\
t^{(n+1)}(x)&=t_0-\frac{1}{mc^{3}}\int_{x_0}^{x}q^{(n+1)}(\eta)\,d\eta.
\end{align}
To first order in weak spatial dependence (small $\partial_x V_{\rm car}$),
\begin{equation}
t(x)\approx t_0-\frac{q_0}{mc^{3}}(x-x_0)
-\frac{1}{mc^{3}}\int_{x_0}^{x}\!\!\Bigg[\int_{x_0}^{\eta}
\partial_x V_{\rm car}\!\left(\xi,\,t_0-\frac{q_0}{mc^{3}}(\xi-x_0)\right)d\xi\Bigg]d\eta.
\label{eq:first_order_t}
\end{equation}

\paragraph{Exactly solvable cases.}
\emph{(A) Time-only potential $V_{\rm car}(t)$.}
Here $\partial_x V_{\rm car}=0$, hence $q(x)\equiv q_0$, and
\begin{equation}
t(x)=t_0-\frac{q_0}{mc^{3}}(x-x_0),\qquad
p_x=\,\frac{q_0^{2}}{2mc^{3}}.
\label{eq:t_timeonly}
\end{equation}

\emph{(B) Space-only potential $V_{\rm car}(x)$.}
Then $p_t(x)\equiv p_{t,0}$ and $q(x)=p_{t,0}+V_{\rm car}(x)$, so
\begin{equation}
t(x)=t_0-\frac{1}{mc^{3}}\int_{x_0}^{x}\big(p_{t,0}+V_{\rm car}(\xi)\big)\,d\xi,\qquad
p_x(x)=\,\frac{\big(p_{t,0}+V_{\rm car}(x)\big)^{2}}{2mc^{3}}.
\label{eq:t_spaceonly}
\end{equation}
Two useful subcases:
\begin{align}
&V_{\rm car}(x)=V_0+\alpha x
\ \Rightarrow\
t(x)=t_0-\frac{p_{t,0}+V_0}{mc^{3}}(x-x_0)-\frac{\alpha}{2mc^{3}}\big(x^{2}-x_0^{2}\big),\\[3pt]
&V_{\rm car}(x)=\tfrac{\kappa}{2}x^{2}
\ \Rightarrow\
t(x)=t_0-\frac{p_{t,0}}{mc^{3}}(x-x_0)-\frac{\kappa}{6mc^{3}}\big(x^{3}-x_0^{3}\big).
\end{align}

\section{Carroll--Schr\"odinger Inner Product and Generators}\label{section6}

We now treat $x$ as the evolution variable and, for each fixed $x$, view the wavefunction as a vector $\Psi_x(t):=\Psi(x,t)$ in the equal-$x$ Hilbert space $\mathcal H:=L^2(\mathbb R_t,dt)$. With this viewpoint the free Carroll--Schr\"odinger equation
\begin{equation}
i\hbar c\,\partial_x\Psi(x,t)\;+\;\frac{\hbar^2}{2mc^2}\,\partial_t^2\Psi(x,t)\;=\;0
\label{eq:carroll-free}
\end{equation}
reads naturally as a Schr\"odinger-type $x$-evolution,
\begin{equation}
i\hbar\,\partial_x\Psi(x,\cdot)=H_x\,\Psi(x,\cdot),
\qquad
H_x:=\frac{\hbar^2}{2mc^{3}}\big(-\partial_t^2\big),
\label{eq:x-evolution}
\end{equation}
so $H_x$ plays the role of the $x$-evolution generator on $\mathcal H$.

Before analyzing the operators, it is helpful to see directly that the $t$-inner product is conserved along $x$. Multiplying \eqref{eq:carroll-free} by $\Psi^*$ and subtracting the conjugate equation, we obtain the continuity law
\begin{equation}
\partial_x\rho_{t}+\partial_t J_{t}=0,
\qquad
\rho_{t}:=|\Psi|^2,
\qquad
J_{t}:=\frac{\hbar}{mc^3}\,\Im\!\big(\Psi^*\partial_t\Psi\big)
=\frac{\hbar}{2i\,mc^3}\!\big(\Psi^*\partial_t\Psi-\Psi\,\partial_t\Psi^*\big).
\label{eq:continuity}
\end{equation}
(Naming $J_t$ and $\rho_t$ emphasizes their role with respect to the time direction and the inner product.) Now, under standard boundary hypotheses in $t$ (e.g.\ $\Psi,\partial_t\Psi\in L^2(\mathbb R)$ with $J_t(x,t)\to0$ as $t\to\pm\infty$, or periodic/quasi-periodic conditions), integration in $t$ gives
\begin{equation}
\frac{d}{dx}\int_{\mathbb R}|\Psi(x,t)|^2\,dt=0
\quad\Longrightarrow\quad
\|\Psi(x,\cdot)\|_{\mathcal H}=\|\Psi(x_0,\cdot)\|_{\mathcal H}\ \ \text{for all }x,x_0,
\label{eq:norm-conserved}
\end{equation}
so the equal-$x$ inner product $\langle\phi,\psi\rangle_t:=\int_{\mathbb R}\phi(t)^*\psi(t)\,dt$ is conserved along the $x$-flow.

We next analyze the operators. To make self-adjointness transparent we pass to frequency space via the \emph{unitary} Fourier transform
\[
(\mathcal F\psi)(\omega):=\frac{1}{\sqrt{2\pi}}\int_{\mathbb R}e^{-i\omega t}\psi(t)\,dt,
\qquad
\mathcal F^{-1}\widetilde\psi(t)=\frac{1}{\sqrt{2\pi}}\int_{\mathbb R}e^{+i\omega t}\widetilde\psi(\omega)\,d\omega.
\]
In this representation, $-\partial_t^2$ becomes multiplication by $\omega^2$, hence it is self-adjoint on $H^2(\mathbb R)$ (and essentially self-adjoint on $\mathcal S(\mathbb R)$). Consequently,
\begin{equation}
H_x=\frac{\hbar^2}{2mc^{3}}(-\partial_t^2)
\quad\text{is self-adjoint on}\quad D(H_x)=H^2(\mathbb R)\subset\mathcal H.
\end{equation}
Now, defining the energy operator \emph{spectrally}, set
\begin{equation}
(\widehat E\,\widetilde\psi)(\omega):=\hbar\,\omega\,\widetilde\psi(\omega),
\qquad
D(\widehat E)=\Big\{\widetilde\psi\in L^2(\mathbb R_\omega):\ \omega\,\widetilde\psi\in L^2(\mathbb R_\omega)\Big\}.
\label{eq:spectral-E}
\end{equation}
Pulling back by the unitary Fourier transform, we obtain on the time side
\begin{equation}
\boxed{\ \widehat E=\mathcal F^{-1}(\hbar\,\omega)\mathcal F=-\,i\hbar\,\partial_t
\ \ \text{with domain}\ \ D(\widehat E)=H^1(\mathbb R)\ },
\label{eq:energy-operator}
\end{equation}
and, under the $t$-inner product, $\widehat E$ is Hermitian: for $\phi,\psi\in\mathcal S(\mathbb R)$ (or in $H^1$ with the boundary hypotheses above),
\begin{equation}
\langle\phi,\widehat E\psi\rangle_t-\langle\widehat E\phi,\psi\rangle_t
=-\,i\hbar\!\int_{\mathbb R}\!\phi^*(t)\,\partial_t\psi(t)\,dt
+\,i\hbar\!\int_{\mathbb R}\!(\partial_t\phi)^*(t)\,\psi(t)\,dt
=-\,i\hbar\,[\phi^*(t)\psi(t)]_{-\infty}^{+\infty}=0.
\label{eq:E-symmetric}
\end{equation}
Hence $\widehat E$ is self-adjoint on $H^1(\mathbb R)$, and, spectrally, the $x$-generator is simply the quadratic function of $\widehat E$,
\begin{equation}
\boxed{\ H_x=\frac{\widehat E^{\,2}}{2mc^{3}}\ }.
\end{equation}
By Stone’s theorem the family $U_x(a):=\exp\!\big(-\tfrac{i}{\hbar}aH_x\big)$ is a strongly continuous unitary group on $\mathcal H$, and the $x$–evolution is given by
\begin{equation}
\Psi(x,\cdot)=U_x(x-x_0)\,\Psi(x_0,\cdot),
\end{equation}
which is another way to read the conservation law \eqref{eq:norm-conserved}.

Finally, to connect with the Hamilton--Jacobi discussion, we recall our momentum convention $\widehat p:=+\,i\hbar\,\partial_x$, which in WKB implies $p_x=-\partial_x S$. On \emph{solutions} of \eqref{eq:x-evolution} the differential generator coincides with $H_x$, namely $\widehat p\,\Psi=H_x\,\Psi$, and for sufficiently regular solutions $\Phi,\Psi$ we therefore have
\[
\langle\Phi,\widehat p\,\Psi\rangle_t-\langle\widehat p\,\Phi,\Psi\rangle_t
=\langle\Phi,H_x\Psi\rangle_t-\langle H_x\Phi,\Psi\rangle_t=0,
\]
which is nothing but the bilinear version of the continuity law \eqref{eq:continuity}. On the full line $\mathbb R_t$ no endpoint conditions are required for self-adjointness: $-i\hbar\,\partial_t$ is essentially self-adjoint on $C_c^\infty(\mathbb R)$ with closure $H^1(\mathbb R)$, and $H_x$ is self-adjoint on $H^2(\mathbb R)$. On a finite time interval $[T_1,T_2]$, periodic/quasi-periodic or suitable separated boundary conditions ensure the vanishing of the $t$-boundary contribution in \eqref{eq:E-symmetric} and \eqref{eq:continuity}, after which the same argument carries through unchanged.

\section{Solutions Analysis}\label{section7}

\subsection*{Gaussian wave packet}

Consider the free Carroll--Schr\"odinger evolution
\begin{equation}
i\hbar c\,\partial_x \Psi(x,t) + \frac{\hbar^2}{2mc^2}\,\partial_t^2 \Psi(x,t)=0,
\label{eq:CS-Gauss:PDE}
\end{equation}
viewed as a Schr\"odinger-type equation in the evolution variable $x$ on the Hilbert space $\mathcal H=L^2(\mathbb R_t,dt)$, with self-adjoint generator
\begin{equation}
i\hbar\,\partial_x\Psi=H_x\Psi,
\qquad
H_x=\frac{\hbar^2}{2mc^3}\,(-\partial_t^2)
\quad\text{on } H^2(\mathbb R_t).
\end{equation}
We take a normalized Gaussian in time at $x=0$,
\begin{equation}
\Psi(0,t)=\Psi_0(t):=\frac{1}{(\pi\sigma^2)^{1/4}}\,
\exp\!\left[-\frac{(t-t_0)^2}{2\sigma^2}\right],
\qquad \|\Psi_0\|_{L^2(\mathbb R_t)}=1,
\label{eq:CS-Gauss:IC}
\end{equation}
with temporal width $\sigma>0$ and center $t_0\in\mathbb R$.

Using the unitary Fourier transform in $t$,
\[
\widetilde\Psi(x,\omega)=\frac{1}{\sqrt{2\pi}}\!\int_{\mathbb R}e^{-i\omega t}\Psi(x,t)\,dt,
\qquad
\Psi(x,t)=\frac{1}{\sqrt{2\pi}}\!\int_{\mathbb R}e^{+i\omega t}\widetilde\Psi(x,\omega)\,d\omega,
\]
we have $\partial_t^2\!\to -\omega^2$, hence
\[
\partial_x \widetilde\Psi(x,\omega)
=-\,i\,\beta\,\omega^2\,\widetilde\Psi(x,\omega),
\qquad
\beta:=\frac{\hbar}{2mc^3},
\]
so $x$–evolution multiplies each frequency component by the pure phase $e^{-i\beta x\omega^2}$. Evaluating the Gaussian integral gives the exact solution
\begin{equation}
\Psi(x,t)
= \frac{1}{(\pi\sigma^2)^{1/4}}\,
\frac{1}{\sqrt{D(x)}}\,
\exp\!\left[-\,\frac{(t-t_0)^2}{2\sigma^2\,D(x)}\right],
\qquad
D(x):=1 + i\,\chi(x),\quad
\chi(x):=\frac{\hbar x}{m c^3\,\sigma^2}.
\label{eq:CS-Gauss:solution}
\end{equation}

The conserved density and current (from $\partial_x\rho_t+\partial_t J_t=0$, cf.\ Eq.~\ref{eq:continuity}) are
\begin{equation}
\rho_t(x,t)=|\Psi(x,t)|^2
=\frac{1}{\sqrt{\pi}\,\sigma\,\sqrt{1+\chi(x)^2}}\,
\exp\!\left[-\,\frac{(t-t_0)^2}{\sigma^2\bigl(1+\chi(x)^2\bigr)}\right],
\label{eq:CS-Gauss:rho}
\end{equation}
\begin{equation}
J_t(x,t)
=\frac{\hbar}{mc^3}\,\Im\!\big(\Psi^*\partial_t\Psi\big)
=\frac{\hbar}{mc^3}\,
\frac{\chi(x)}{\sigma^2\bigl(1+\chi(x)^2\bigr)}\,(t-t_0)\,\rho_t(x,t).
\label{eq:CS-Gauss:J}
\end{equation}
In particular, the equal-$x$ norm is preserved for all $x$; explicitly,
\begin{equation}
\int_{\mathbb R}\rho_t(x,t)\,dt
=\frac{1}{\sqrt{\pi}\,\sigma\sqrt{1+\chi(x)^2}}
\int_{\mathbb R}\exp\!\left[-\frac{(t-t_0)^2}{\sigma^2\bigl(1+\chi(x)^2\bigr)}\right] dt
=1,
\label{eq:CS-Gauss:norm}
\end{equation}
by $\int_{\mathbb R}e^{-y^2/a^2}dy=\sqrt{\pi}\,a$. Physically, this parallels the usual Schr\"odinger packet with the roles of space and time exchanged: the packet remains centered at $t_0$ while its temporal width disperses as
\begin{equation}
\sigma_{\mathrm{eff}}(x)=\sigma\,\sqrt{1+\chi(x)^2}
=\sqrt{\sigma^2 + \left(\frac{\hbar x}{m c^3\,\sigma}\right)^2},
\label{eq:CS-Gauss:width}
\end{equation}
and $J_t(x,t)$ is odd about $t_0$, encoding the $t$-flux that balances the $x$-rate of change of $\rho_t$ in the continuity equation. Since $H_x$ is self-adjoint on $H^2(\mathbb R_t)$, the $x$–evolution is unitary on $L^2(\mathbb R_t)$, which implies \eqref{eq:CS-Gauss:norm} under the boundary conditions stated in Section~\ref{section6}.

Finally, if one adds a carrier phase to the initial datum,
\[
\Psi_0(t)\ \longrightarrow\ \Psi_0(t)\,e^{-i\omega_0(t-t_0)}
\quad (E_0=\hbar\omega_0),
\]
the solution acquires the same complex width $D(x)$ but its center drifts. A stationary-phase estimate of the inverse-transform phase $\omega t-\beta x\omega^2$ gives
\begin{equation}
t_{\mathrm c}(x)=t_0+\frac{E_0}{mc^3}\,x,
\qquad\Rightarrow\qquad
\frac{dt_{\mathrm c}}{dx}=\frac{E_0}{mc^3}
=\frac{d\kappa}{d\omega}\bigg|_{\omega_0},
\end{equation}
since the $x$–wavenumber is $\kappa(\omega)=\beta\omega^2$. Thus $dx/dt_{\mathrm c}=mc^3/E_0$ equals the classical (ray) velocity obtained from the Hamilton--Jacobi/characteristics analysis \eqref{exoticmomentum}, in perfect agreement with the dispersion picture.

\subsection*{Finite–time perturbations for a purely time–dependent potential}
\label{subsec:time-only-potential}

We consider the $(1+1)$–dimensional equation
\begin{equation}
  i\hbar c\,\partial_x \Psi(x,t)
  -\frac{1}{2 m c^{2}}\!\left(-\,i\hbar\,\partial_t - V(t)\right)^{\!2}\Psi(x,t)=0,
  \label{eq:master}
\end{equation}
for a scalar wavefunction $\Psi(x,t)$ coupled to a time–dependent scalar potential $V(t)$. 
Seeking separable solutions of the form
\begin{equation}
  \Psi(x,t)=\psi_t(t)\,\psi_x(x),
\end{equation}
Eq.~\eqref{eq:master} yields
\begin{equation}
  i\hbar c\,\frac{\psi_x'(x)}{\psi_x(x)}
  = \frac{1}{2 m c^{2}}\,
    \frac{\left(i\hbar\,\frac{d}{dt}+V(t)\right)^{2}\psi_t(t)}{\psi_t(t)}
  = cp_0,
  \label{eq:separation}
\end{equation}
The spatial factor solves
$i\hbar c\,\psi_x'(x)=E_0^{2}\psi_x(x)$ and is therefore a plane wave,
\begin{equation}
  \psi_x(x)=\exp\!\left(\frac{i p_0 x}{\hbar}\right), 
  \qquad \text{with } \; E_0^{2} \equiv 2 m c^{3} p_0 ,
  \label{eq:plane-wave}
\end{equation}
for some real constant $p_0$ fixed by boundary conditions. The temporal factor obeys
\begin{equation}
  \left(i\hbar\,\frac{d}{dt}+V(t)\right)^{2}\psi_t(t)=E_0^{2}\,\psi_t(t).
  \label{eq:time-eq}
\end{equation}

A convenient gauge transformation removes the explicit potential from
\eqref{eq:time-eq}. Define
\begin{equation}
  \phi(t):=\exp\!\left(\frac{i}{\hbar}\int_{t_0}^{t} V(t')\,dt'\right)\psi_t(t).
  \label{eq:gauge}
\end{equation}
Using \eqref{eq:gauge} in \eqref{eq:time-eq} gives the free second–order equation
\begin{equation}
  \frac{d^{2}\phi}{dt^{2}}+\frac{E_0^{2}}{\hbar^{2}}\,\phi(t)=0.
  \label{eq:phi-ode}
\end{equation}

Now, for illustrative purposes, suppose the perturbation acts only on a finite interval $t\in[0,T]$. Consistency with compact support in time enforces the homogeneous boundary conditions
\begin{equation}
  \phi(0)=\phi(T)=0.
  \label{eq:bc}
\end{equation}
Equations \eqref{eq:phi-ode}–\eqref{eq:bc} admit the discrete family of solutions
\begin{equation}
  \phi_n(t)=A\,\sin\!\left(\frac{E_n t}{\hbar}\right), 
  \qquad
  E_n=\frac{n\pi\hbar}{T}, \quad n\in\mathbb{N}.
  \label{eq:phi-spectrum}
\end{equation}
Thus, restricting the interaction to a finite temporal interval leads to an energy quantization
$E_n=n\pi\hbar/T$, in direct analogy with the spatial quantization for a particle in a finite box in the Schr\"odinger problem.

Imposing unit normalization in time over the active window,
\begin{equation}
  \int_{0}^{T}\! |\Psi(x,t)|^{2}\,dt = 1,
  \label{eq:time-norm}
\end{equation}
fixes the amplitude in \eqref{eq:phi-spectrum} to
\begin{equation}
  A=\sqrt{\frac{2}{T}}.
  \label{eq:A}
\end{equation}
Combining \eqref{eq:gauge}, \eqref{eq:plane-wave}, and \eqref{eq:A}, the normalized modes read
\begin{equation}
  \Psi_{n}(x,t)
  = \exp\!\left(\frac{i p_0 x}{\hbar}\right)
    \exp\!\left(\frac{i}{\hbar}\int_{t_0}^{t} V(t')\,dt'\right)
    \sqrt{\frac{2}{T}}\,
    \sin\!\left(\frac{n\pi t}{T}\right),
  \qquad 0\le t\le T .
  \label{eq:psi-full}
\end{equation}
Because the gauge factor in \eqref{eq:psi-full} has unit modulus, the probability density is independent of $V(t)$,
\begin{equation}
  \rho_{n}(x,t):=|\Psi_{n}(x,t)|^{2}=\frac{2}{T}\,\sin^{2}\!\left(\frac{n\pi t}{T}\right).
  \label{eq:density}
\end{equation}
Analyzing the temporal current, for the present spatially stationary state, the temporal component of the conserved current reduces to two equal and opposite contributions proportional to $V(t)$, namely
\begin{equation}
  J_{t}(t) 
  = -\,\frac{V(t)}{m c^{2}}\,\rho_{n}(x,t)
    + \frac{V(t)}{m c^{2}}\,\rho_{n}(x,t)
  = 0,
  \label{eq:Jt}
\end{equation}
so that the temporal current vanishes identically while the density \eqref{eq:density} remains unaffected by the detailed time profile of the potential.

\subsection*{General space–time dependent potential}
\label{subsec:general-potential-regular}

We now allow the interaction to depend on both space and time,
$V=V(x,t)$, and consider the equation
\begin{equation}
  i\hbar c\,\partial_x \Psi(x,t)
  -\frac{1}{2 m c^{2}}\!\left(-\,i\hbar\,\partial_t - V(x,t)\right)^{\!2}\Psi(x,t)=0 .
  \label{eq:general-master}
\end{equation}
Introducing the gauge transform
\begin{equation}
  \phi(x,t)
  := \exp\!\left(\frac{i}{\hbar}\int_{t_0}^{t} V(x,\tau)\,d\tau\right)\Psi(x,t) ,
  \label{eq:general-gauge}
\end{equation} 
a direct computation shows that the potential cancels from the second-order
time operator, while the spatial derivative acquires a memory term:
\[
\partial_t\phi
=\exp\!\left(\tfrac{i}{\hbar}\!\int_{t_0}^{t}\!V\,d\tau\right)\partial_t\Psi
+\frac{i}{\hbar}V(x,t)\,\phi, 
\quad
\partial_x\phi
=\exp\!\left(\tfrac{i}{\hbar}\!\int_{t_0}^{t}\!V\,d\tau\right)\partial_x\Psi
+\frac{i}{\hbar}\Bigg(\int_{t_0}^{t}\!\partial_x V(x,\tau)\,d\tau\Bigg)\phi .
\]
Substituting into \eqref{eq:general-master} yields the reduced equation
\begin{equation}
  i\hbar c\,\partial_x \phi(x,t)
  + \frac{\hbar^{2}}{2 m c^{2}}\,\partial_{t}^{2}\phi(x,t)
  - c\Bigg(\int_{t_0}^{t}\partial_x V(x,\tau)\,d\tau\Bigg)\phi(x,t)=0 .
  \label{eq:reduced-general}
\end{equation}
Equivalently, isolating the $x$-evolution gives a Schr\"odinger--type equation
with $x$ playing the role of ``time'',
\begin{equation}
  i\hbar c\,\partial_x \phi(x,t)
  = -\,\frac{\hbar^{2}}{2 m c^{2}}\,\partial_{t}^{2}\phi(x,t)
    + c\,F(x,t)\,\phi(x,t),
  \qquad
  F(x,t):=\int_{t_0}^{t}\partial_x V(x,\tau)\,d\tau .
  \label{eq:schrodinger-in-x}
\end{equation}
Since the transformation \eqref{eq:general-gauge} is unitary, one has
$|\phi|^{2}=|\Psi|^{2}$.

Analyzing Eq.~\ref{eq:schrodinger-in-x}, we note that the gauge-transformed equation can also be obtained by viewing the interaction as entering through the momentum. In particular, to include an interacting case (analogous to a scalar potential in the Schr\"odinger equation), it is equivalent to implement a minimal substitution on the momentum,
\begin{equation}
  p \;\longmapsto\; p' = p - F(x,t),
  \label{eq:minimal-coupling-momentum}
\end{equation}
where $F(x,t)$ denotes the interaction momentum, in direct analogy with the standard energy shift $E\mapsto E' = E - V(x,t)$ in
nonrelativistic quantum mechanics. This is consistent with the Carrollian
viewpoint in which dynamics is governed by the
momentum sector rather than by the energy.

It is worth noting that Eq.~\ref{eq:schrodinger-in-x} has a precise structural similarity to the equations describing the evolution of temporal solitons \cite{AgrawalNFO2019, ZhengLiu2022}. This is important because it suggests that temporal solitons can be regarded as inherently exhibiting Carroll symmetry.

Now, in order to solve Eq.~\ref{eq:schrodinger-in-x} for a general space–time interaction,
we proceed with perturbation theory, splitting the potential into a solvable time–dependent part plus a small
space–dependent perturbation:
\begin{equation}
  c\,f(x,t) \;=\; g(t) \;+\; \varepsilon\,\eta(x),
  \qquad 0<\varepsilon\ll 1 .
  \label{eq:split-potential}
\end{equation}
Define the unperturbed $x$-evolution operator $\mathcal{U}_{0}(x,x_{0})$
through
\begin{equation}
  i\hbar c\,\partial_x \mathcal{U}_{0}(x,x_{0})
  = \Big[-\,\tfrac{\hbar^{2}}{2 m c^{2}}\,\partial_{t}^{2} - g(t)\Big]
    \mathcal{U}_{0}(x,x_{0}),
  \qquad \mathcal{U}_{0}(x_{0},x_{0})=\mathbf{1}.
  \label{eq:U0}
\end{equation}
(When $g(t)$ alone is present, the modes coincide with those constructed in
Sec.~\ref{subsec:time-only-potential} after the unitary gauge reduction.)
The full solution admits a Dyson expansion in the ``spatial time'' $x$:
\begin{equation}
  \phi(x,t)
  = \mathcal{U}_{0}(x,x_{0})\,\phi(x_{0},t)
    -\frac{i\,\varepsilon}{\hbar c}\int_{x_{0}}^{x}
    \mathcal{U}_{0}(x,\xi)\,\eta(\xi)\,\mathcal{U}_{0}(\xi,x_{0})\,
    \phi(x_{0},t)\,d\xi \;+\; \mathcal{O}(\varepsilon^{2}).
  \label{eq:first-order-solution}
\end{equation}
Equations \eqref{eq:schrodinger-in-x}–\eqref{eq:first-order-solution}
provide a consistent and systematic framework to treat general interactions
$V(x,t)$: the gauge transform removes the explicit $V$ from the temporal
operator, converts the problem into Schr\"odinger--type evolution in $x$, and
relates the usual scalar potential to an \emph{interaction momentum}
$F(x,t)$ via \eqref{eq:minimal-coupling-momentum}.

\section{Discussion}
The framework developed here in $1{+}1$ dimensions suggests several natural extensions. A first direction is to test which parts of the operator–level dictionary survive in higher dimensions. In particular, one may ask whether the potential–dependent map $x=\delta(t)$ admits a meaningful multidimensional analogue. Beyond kinematics, one should revisit the equal–$x$ Hilbert–space picture: the natural choice $L^{2}(\mathbb{R}^{d}_{t},dt)$ remains available under the assumption that there still exists a single time dimension for higher–dimensional generalizations.

There already exist nontrivial $3$–dimensional generalizations of the Carroll--Schrödinger equation, with distinct structural choices and symmetry inputs; exploring our derivations against those setups would help isolate which features are universal. For instance, recent proposals for three–dimensional Carrollian quantum dynamics~\cite{Najafizadeh2025,CasanovaRojasArias2025} suggest alternative expressions for the generalized 3D equation; it is important to mention that both of these generalizations treat time as still 1–dimensional and space as 3–dimensional. It would be natural to check (i) the behavior of Carroll particles in the classical Hamilton--Jacobi limit and how they differ from classical Newtonian dynamics, (ii) how the interaction–momentum picture extends to a vector field $\,\mathbf{f}(x,t)$ (and its physical interpretation), (iii) whether the gauge/inversion map relating continuity equations lifts to a tensorial statement for the full Carroll current in higher rank, and (iv) whether there exists an expansion for Klein--Gordon solutions in terms of Schr\"odinger and Carroll--Schrödinger solutions.

Finally, exploring quantum field theory interactions, many–body extensions (both bosonic and fermionic), and the inclusion of external electromagnetic backgrounds provide fertile ground to explore Carrollian solutions and their implications.

\section{Conclusions}
We presented a unified operator framework relating the Schrödinger and Carroll--Schrödinger equations in $1{+}1$ dimensions. The main points are: (i) a precise shared–solutions criterion $[\hat{\mathcal H},\hat{\mathcal F}]=0$ that fixes compatible potentials and yields an explicit common solution class, together with a stricter “identical separated solutions’’ option; (ii) a potential–dependent reparametrization $x=\delta(t)$ mapping the space–independent Carroll problem to the time–independent Schrödinger problem, an eliminant leading to a Schwarzian master relation, and explicit reconstructions for harmonic, Coulomb–like, and free targets; (iii) an identification of Carrollian and Schrödinger continuity equations via a gauge removal and a pseudo–coordinate inversion, plus an equal–$x$ Hilbert–space formulation with unitary $x$–evolution; (iv) exact solutions including a closed–form Gaussian packet in $t$ whose dispersion law matches the Schrödinger Gaussian under the gauge/inversion map and agrees in the Hamilton--Jacobi (semiclassical) limit, as well as finite–time quantization for time–localized interactions; (v) a gauge reduction for general $V(x,t)$ yielding an \emph{interaction momentum} and a controlled Dyson expansion in the spatial evolution variable with direct physical analogies.

These results provide a practical dictionary for translating potentials, solutions, and conserved structures between the two formalisms. The discussion above outlines several next steps: extending the construction to $3{+}1$ dimensions in light of recent Carroll--Schrödinger generalizations~\cite{Najafizadeh2025,CasanovaRojasArias2025}, including background fields and many–body sectors, and generalizing the Schwarzian potential inversion procedure and classical limit of Carroll--Schrödinger dynamics from the quantum equation.

\end{document}